%Paper: hep-th/9306016
%From: Palev Tchavpar <palev@ictp.trieste.it>
%Date: Thu, 3 Jun 93 10:36:41 MET DST

\baselineskip 18pt
\leftskip 36pt
\noindent
{\bf QUANTISATION OF U$_q$[OSP(1/2N)] WITH DEFORMED PARA-BOSE
\hfil \break OPERATORS}

\vskip 32pt
\noindent
T. D. Palev\footnote*{Permanent address: Institute for Nuclear Research
and Nuclear Energy, 1784 Sofia, Bulgaria; E-mail before July 31, 1993
palev@ictp.trieste.it, after July 31, 1993: palev@bgearn.bitnet}

\noindent
Applied Mathematics and Computer Science, University of Ghent,
B-9000 Gent, Belgium

\noindent
and

\noindent
International Centre for Theoretical Physics, 34100 Trieste, Italy

\vskip 32pt
{\bf Abstract.} The observation that $n$ pairs of para-Bose
(pB) operators generate the universal enveloping algebra of the
orthosymplectic Lie superalgebra $osp(1/2n)$ is used in order to
define deformed pB operators. It is shown that these operators
are an alternative to the Chevalley generators. On this
background $U_q[osp(1/2n)]$, its "Cartan-Weyl" generators and
their "supercommutation" relations are written down entirely in
terms of deformed pB operators. An analog of the Poincar\'e-
Birkhoff-Witt theorem is formulated.

\vskip 48pt

\noindent
Soon after  parastatistics has been invented [1], it was
realized that it carries a deep algebraic structure. More precisely,
any $n$ pairs of para-Fermi operators generate the simple  Lie algebra
$B_n \equiv so(2n+1)  $ [2,3], whereas n pairs of para-Bose
creation and annihilation operators (CAO's)
${\hat A}_1^\pm, \ldots , {\hat A}_n^\pm$ generate
a Lie superalgebra [3], which is
isomorphic  to one of the basic Lie superalgebras (LS's) in
the classification of Kac [4], namely to the orthosymplectic Lie
superalgebra $osp(1/2n) \equiv B(0/n)$ [5]. In somewhat more
implicit form the para-Bose operators (pB operators) were
introduced even earlier by Wigner [6] in a search of the most
general commutation relations between a  position operator $q$
and a momentum operator $p$ of a one dimensional oscillator,
so that the Heisenberg equations are
identical with the Hamiltonian equations.
The operators $p, q$ turned out to generate
$osp(1/2)$ and in fact Wigner was the first to find a
class of (infinite-dimensional) representations of a Lie
superalgebra [7]. Later on the results of Wigner gave rise to more
general quantum systems (see [8] for references in this respect
and for a general introduction to parastatistics), and in
particular to quantum systems related to the classes {\bf A, B,
C, D} of basic LS's [9].

Purely algebraically the pB operators are defined
as operators, which satisfy the relations
($\xi, \eta, \epsilon = \pm$ or $\pm 1$, $i,j,k=1,2,\ldots ,n$ ;
$ [x,y]=xy-yx,\; \{x,y\}=xy+yx$)

$$[\{{\hat A}_i^\xi,{\hat A}_j^\eta \},{\hat A}_k^\epsilon]
=(\epsilon - \xi)\delta_{ik}
{\hat A}_j^\eta + (\epsilon - \eta) \delta_{jk}{\hat A}_i^\xi.
\eqno(1)$$

\noindent
Let $L(n)=lin.env.\big\{\{{\hat A}_i^\xi,{\hat A}_j^\eta \},
{\hat A}_k^\epsilon \mid
\xi, \eta, \epsilon = \pm,i,j,k=1,2,\ldots ,n \big\} $
be a ${\bf Z}_2$ graded linear space with
the pB operators as odd elements. Define
a supercommutator on it as an anticommutator between any two odd
elements and as a commutator otherwise.

{\it Proposition 1 } [4]. $L(n)$ is a LS, isomorphic to
$osp(1/2n)$, with an odd
subspace, spanned by the pB operators, and an even subalgebra
$sp(2n)=lin.env.\big\{\{{\hat A}_i^\xi,{\hat A}_j^\eta \} \mid
\xi, \eta, = \pm,i,j=1,2,\ldots ,n \big\}$. The pB operators
define uniquely $osp(1/2n)$.
The associative superalgebra with unity, the pB operators as
free generators, and the relations (1) is the universal
enveloping algebra $U[osp(1/2n)]$ of $osp(1/2n)$.
The set of all (arbitrarily) ordered monomials of the operators

$$H_i=-{1\over 2}\{{\hat A}_i^-,{\hat A}_i^+ \},
\quad {\hat A}_i^{\pm},\quad
\{{\hat A}_i^-,{\hat A}_j^+ \}
,\quad \{{\hat A}_i^\xi,{\hat A}_j^\xi \},\quad i\neq j=
1,\ldots,n,\quad \xi=\pm, \eqno(2)$$

\noindent
constitute a basis in $U[osp(1/2n)]$, whereas the operators (2)
together with all $({\hat A}_i^\pm)^2$ define a basis, a Cartan-Weyl
basis, in $osp(1/2n)$.

The proof is based on the relations, following
from (1) and (2), namely

$$[H_i,{\hat A}_j^{\pm}]=\mp \delta_{ij}{\hat A}_j^{\pm}, \quad
\{{\hat A}_i^-,{\hat A}_i^+ \}=-2H_i, \eqno(3) $$
$$[\{{\hat A}_i^+,{\hat A}_j^- \},{\hat A}_k^+]=\delta_{jk}
{\hat A}_i^+, \quad
[\{{\hat A}_i^+,{\hat A}_j^- \},{\hat A}_k^-]=\delta_{ik}
{\hat A}_j^-,\quad i\neq j, \eqno(4)  $$
$$[\{{\hat A}_i^\xi,{\hat A}_j^\xi \},{\hat A}_k^\xi]=0,
\quad i\neq j.\eqno(5) $$

\noindent
It is essential to point out that the
eqs.(3)-(5) define uniquely the  supercommutation
relations between all Cartan-Weyl generators. In other words,
$U[osp(1/2n)]$ can be defined as an algebra with free generators
$H_i,\; {\hat A}_i^{\pm} \quad i=1,\ldots,n $ and the relations
(3)-(5).

On the ground of Proposition 1 and using the circumstance that
$U[osp(1/2n)]$ has already been quantized, i.e.,
deformed to a superalgebra $U_q[osp(1/2n)] \equiv U_q$,
which preserves its
Hopf algebra structure [10,11], one can introduce the concept
of deformed pB operators. The deformation of
$U_q$ was carried out
in a Chevalley basis. Clearly, it deforms all elements of
$U_q$ and in particular the pB operators
${\hat A}_1^\pm, \ldots , {\hat A}_n^\pm$. The purpose of the present
letter is to give explicit relations for the deformed pB
operators. We will show that the deformed pB operators
define entirely the superalgebra. We will write down also the
quantum analog of the Cartan Weyl generators in terms of pB
operators and the supercommutation relations between
the Cartan-Weyl generators (2). To our knowledge such relations
for any $n$ have not been given explicitily even in terms of
the Chevalley generators.
So far a deformation of pB operators was carried out for
$n=1$ [12] and $n=2$ [13,14] cases; there have been several other
deformations, which were not endowed with a Hopf algebra
structure [15], [16] ( see also the references therein).

We proceed to introduce $U_q$ .
The Cartan matrix $(\alpha_{ij})$ is chosen
as in [11], i.e., as a $n  \times n $ symmetric matrix with
$\alpha_{nn}=1$, $\alpha_{ii}=2, \; i=1,\ldots,n-1 $,
$\alpha_{j,j+1}=\alpha_{j+1,j}=-1, \; j=1,\ldots,n-1, $
and all other $\alpha_{ij}=0$ . Let
$[x]={ q^{x}-q^{-x}\over q-q^{-1} }$. Then $U_q$
is the free associative superalgebra with Chevalley generators
$e_i,\; f_i,\; k_i=q^{h_i},\; i=1,\ldots,n, $  which satisfy
the Cartan relations

$$k_ik_i^{-1}=k_i^{-1}k_i=1, \quad k_ik_j=k_jk_i, \quad
 i,j=1,\ldots,n, \eqno(6) $$

$$k_ie_j=q^{\alpha_{ij}}e_jk_i, \quad
k_if_j=q^{-\alpha_{ij}}f_jk_i, \quad  i,j=1,\ldots,n, \eqno(7)  $$

$$\{ e_n,f_n\} ={{k_n-k_n^{-1}}\over{q-q^{-1}}}, \quad
[e_i,f_j]=\delta_{ij}{{k_i-k_i^{-1}}\over{q-q^{-1}}}=
\delta_{ij}[h_i]
\quad \forall \; i,j=1,\ldots,n \; \; {\rm except} \;i=j=n,
\eqno(8) $$

\noindent
the Serre relations for the simple positive root vectors

$$ [e_i,e_j]=0, \quad if \quad i,j=1,\ldots,n \quad
and \quad \vert i-j \vert >1,
\eqno(9)  $$

$$ e_i^2e_{i+1}-(q+q^{-1})e_ie_{i+1}e_i+e_{i+1}e_i^2=0,
   \quad i=1,\ldots,n-1, \eqno(10)  $$

$$ e_i^2e_{i-1}-(q  +q^{-1})e_ie_{i-1}e_i+e_{i-1}e_i^2=0,
   \quad i=2,\ldots,n-1, \eqno(11)  $$

$$e_n^3e_{n-1}+(1-q  -q^{-1})(e_n^2e_{n-1}e_n+e_ne_{n-1}e_n^2)+
e_{n-1}e_n^3=0, \eqno(12)  $$

\noindent
and the Serre relations obtained from (9)-(12) by replacing
everywhere $e_i$ by $f_i$. The grading on $U_q$ is induced from
the requirement that the generators $e_n,\; f_n$ are odd and all
other generators are even. Throughout
$[a,b]_{q^n}=ab-(-1)^{deg(a)deg(b)}q^nba $
and it is assumed that the deformation parameter $q$ is any
complex number except $q=0$, $q=1$ and $q^2=1$.
The eqs. (6)-(12) are invariant with respect to the antiinvolution
$(e_i)^*=f_i$, $(k_i)^*=k_i^{-1}$, $(q)^*=q^{-1}$.

The action of the coproduct $\Delta$,
the antipode $S$ and the counit $\varepsilon $ reads [11]:

$$\Delta (e_i)=e_i \otimes 1 + k_i \otimes e_i,\quad
\Delta (f_i)=f_i \otimes k_i^{-1} + 1 \otimes f_i ,\quad
\Delta(k_i)=k_i\otimes k_i, \eqno(13)   $$

$$S(e_i)=-k_i^{-1}e_i, \quad
S(f_i)=-f_ik_i, \quad
S(k_i)=k_i^{-1}, \eqno(14)   $$

$$\varepsilon(e_i)= \varepsilon(f_i) = \varepsilon(k_i) =0,
\quad \varepsilon(1)=1. \eqno(15)$$

Define  $2n$ odd elements from
$U_q[osp(1/2n)]$ as  $(i=1,\ldots, n-1)$

$$\vcenter{\openup3\jot\halign {$#$ \hfil \cr
A_i^-=-\sqrt{2}
[e_i,[e_{i+1},[e_{i+2},[\ldots,[e_{n-2},[e_{n-1},e_n]_{q^{-1}}
]_{q^{-1}}\ldots ]_{q^{-1}}, \cr
A_n^-=-\sqrt{2}e_n \cr
}} \eqno(16)   $$
\vskip 12pt

$$\vcenter{\openup3\jot\halign {$#$ \hfil \cr
A_i^+=\sqrt{2}
[\ldots,[f_n,f_{n-1}]_{q},f_{n-2}]_{q},\ldots]_{q},
f_{i+2}]_{q},f_{i+1}]_{q},f_{i}]_{q}, \cr
A_n^+=\sqrt{2}f_n, \cr
}} \eqno(17)   $$

\noindent
and another $n$ even "Cartan" elements

$$K_i=k_ik_{i+1} \ldots k_n=q^{H_i}, \quad H_i=h_i + \ldots + h_n
\quad i=1,\ldots,n. \eqno(18)  $$

We call the operators
$ K_i^{\pm 1}, \;A_i^\pm, \quad i=1,\ldots ,n $
pre-oscillator generators, since
in a certain representation of $U_q[osp(1/2n)]$ [17] these
operators generate the oscillator algebra of the
deformed Bose creation and annihilation operators as introduced
in [18-20] or in [21] .  The antiinvolution on them reads:
$(A_i^{\pm})^*=-A_i^{\mp}, \; (K_i)^*=K_i^{-1},\quad i=1,\ldots,n.$

A substantial part of all computational time went to
derive the following relations:

$$K_iA_j^{\pm}=q^{\mp \delta_{ij}}A_j^{\pm}K_i,
\quad \{A_i^-,A_i^+ \}=-2{{K_i-K_i^{-1} \over {q-q^{-1}}}}
\equiv -2[H_i], \quad i=1,\ldots ,n, \eqno(19)   $$

$$\vcenter{\openup3\jot \halign{$#$ \hfil & \hskip 12pt $#$ \hfil
 \cr
[\{A_i^-,A_j^+ \},A_k^+]=0,
&[\{A_i^-,A_j^+ \},A_j^+]_{q^{-1}}=0, \quad if \quad i<j<k \quad
 or \quad k<i<j,\cr
[\{A_i^-,A_j^+ \},A_i^+]=2K_iA_j^+,
& [\{A_i^-,A_j^+ \},A_k^+]=(q^{-1}-q)\{A_i^-,A_k^+ \}A_j^+,
\quad if \quad i<k<j,\cr
[\{A_i^-,A_j^+ \},A_k^-]=0,
&[\{A_i^-,A_j^+ \},A_i^-]_{q}=0, \quad if \quad k<i<j \quad
 or \quad i<j<k,\cr
[\{A_i^-,A_j^+ \},A_j^-]=-2K_jA_i^-,
& [\{A_i^-,A_j^+ \},A_k^-]=(q-q^{-1})\{A_k^-,A_j^+ \}A_i^-,
\quad if \quad i<k<j,\cr
[\{A_i^-,A_j^+ \},A_k^-]=0,
&[\{A_i^-,A_j^+ \},A_i^-]_{q^{-1}}=0, \quad if \quad j<i<k \quad
 or \quad k<j<i,\cr
[\{A_i^-,A_j^+ \},A_j^-]=-2K_j^{-1}A_i^-,
& [\{A_i^-,A_j^+ \},A_k^-]=(q^{-1}-q)\{A_k^-,A_j^+ \}A_i^-,
\quad if \quad j<k<i,\cr
[\{A_i^-,A_j^+ \},A_k^+]=0,
&[\{A_i^-,A_j^+ \},A_j^+]_{q}=0, \quad if \quad k<j<i \quad
 or \quad j<i<k,\cr
[\{A_i^-,A_j^+ \},A_i^+]= 2K_i^{-1}A_j^-,
& [\{A_i^-,A_j^+ \},A_k^+]=(q-q^{-1})\{A_i^-,A_k^+ \}A_j^+,
\quad if \quad j<k<i,\cr
}} \eqno(20)$$

$$\vcenter{\openup3\jot \halign{$#$ \hfil & \hskip 18pt $#$ \hfil
 \cr
[\{A_i^\xi,A_j^\xi \},A_k^\xi]_{q^2}=0,
&[\{A_i^\xi,A_j^\xi \},A_j^\xi]_{q}=0, \quad if \quad i<j<k
\quad \xi=\pm  \cr
[\{A_i^\xi,A_j^\xi \},A_k^\xi]_{q^{-2}}=0,
&[\{A_i^\xi,A_j^\xi \},A_j^\xi]_{q^{-1}}=0, \quad if \quad k<i<j
\quad \xi=\pm  \cr
[\{A_i^\xi,A_j^\xi \},A_k^\xi]=0, & if \quad i<k<j,\quad \xi=\pm.  \cr
}} \eqno(21)$$
\noindent
To this end the following lemma turned to be particularly useful.

{\it Lemma 1.} If $[a,b]=0$, then $[a,[c,b]_q]_q=[[a,c]_q,b]_q$,
$[a,[c,b]_{q^{-1}}]_q=[[a,c]_q,b]_{q^{-1}}$; if $a$, $b$
are even elements and
$[a,c]=0$,
then $(q+q^{-1})[b,[a,[b,c]_q]_q]=[a,[b,[b,c]_q]_{q^{-1}}]_{q^2}-
[[b,[b,a]_q]_{q^{-1}},c]_{q^2}$.

In particular, if $[a_i,a_j]=0$ for all $\vert i-j \vert>1$, then
$$[a_k,[a_{k-1},[a_{k-2}[ \ldots ,[a_3,[a_2,a_1]_q \ldots ]_q=
[\ldots[a_k,a_{k-1}]_q ,a_{k-2}]_q,\ldots
]_q,a_3]_q,a_2]_q,a_1]_q. \eqno(22)$$

{\it Proposition 2.}
(A) The free algebra ${\hat U}_q$ of the pre-oscillator
generators and the relations (19)-(21) is the quantized
$U_q[osp(1/2n)]$ algebra with
generators $e_i,\; f_i,\; k_i=q^{h_i},\; i=1,\ldots,n, $ and
relations (6)-(12);
(B) The operators
$$ K_i^{\pm 1}, \;A_i^\pm,\;
\{A_i^-,A_j^+ \},\; \{A_i^\xi,A_j^\xi \},\quad i\neq j
=1,\ldots,n \eqno(23)$$
are the analog of the Cartan-Weyl generators (2);
(C) The relations (19)-(21) are the analog of the supercommutation
relations (3), (4), (5) among all Cartan-Weyl generators (2);
(D) The set of all normally ordered monomials [11] of the
Cartan-Weyl generators  (23) constitute a basis in
$U_q[osp(1/2n)]$ (Poincar\'e-Birkhoff-Witt theorem).

We sketch the proof. By construction ${\hat U}_q$ is a subalgebra of
$U_q[osp(1/2n)]$. The expressions of the Chevalley generators
in terms of the pre-oscillator generators read ($i=1,\ldots,n-1$):

$$e_n=-(2)^{-1/2}A_n^-,\; f_n=(2)^{-1/2}A_n^+,\;
e_i={q\over 2}\{A_i^-,A_{i+1}^+\}K_{i+1}^{-1},\;
f_i={1\over 2q}\{A_i^+,A_{i+1}^-\}K_{i+1} .
\eqno(24)  $$

\noindent
{}From (24) and (19)-(21) one derives all relations (6)-(12) among
the Chevalley generator. Hence $U_q[osp(1/2n)]$ is a subalgebra
of ${\hat U}_q$ and therefore $U_q[osp(1/2n)]={\hat U}_q$. This
proves (A). From (16)-(18) one concludes that in the limit
$q \rightarrow 1$ the relations (19), (20), (21) of the
operators (23) reduce to the
relations (3), (4), (5) of the Cartan-Weyl generators (2), which
proves (B) and (C). The normal order between the groups of the
positive root vectors (p.r.v.) $A_i,\{A_i^-,A_j^{\pm} \},\; i<j
$, the negative root vectors (n.r.v.) and the Cartan generators
$K_i$ is arbitrary but fixed. For instance,
p.r.v.$<$n.r.v.$<K_i$. The order among the p.r.v. can be taken to
be

$\{A_i^-,A_p^+ \}<A_i^-<\{A_i^-,A_q^- \}<
\{A_j^-,A_r^+ \}<A_j^-<\{A_j^-,A_s^- \}  $

\noindent
for all $i<p, q; \quad i<j ;\quad j<r,s$ and similarly for the
n.r.v. The possibility for such ordering and hence the proof
of (D) follows directly from the relations (19)-(21).

{\it Remark.} The relations (19)-(21) are by no means the
minimal set of relation among the pre-oscillator generators,
which define $U_q[osp(1/2n)]$. Just on the contrary - they
are analog of all supercommutation relations among the
generators in the nondeformed case.\footnote*{After the work
on the present investigation has been completed, we have received
a preprint from L. Hadjiivanov [22], where the question about
a minimal set of relations of the pB operators, defining
$U_q[osp(1/2n)]$, is settled.}

We observe that the Cartan-Weyl generators are expressed in an
easy way in terms of the pre-oscillator generators and even the
anticommutators in (23) remain undeformed. This is due to the
fact that the roots of $A_1^+,\ldots,A_n^+$ are orthogonal to
each other . Certainly, from (16)-(18) and (23)  one can
write down all Cartan-Weyl generators  also in terms of the
Chevalley generators, but the expressions are not so simple.
Here are the Cartan-Weyl generators of the Hopf subalgebra
$U_q[gl(n)]$:

$$\{A_i^-,A_j^+ \}=-2
[\ldots,[f_{i-1},f_{i-2}]_{q},f_{i-3}]_{q},\ldots]_{q},
f_{j+2}]_{q},f_{j+1}]_{q},f_{j}]_{q}
k_i^{-1}k_{i+1}^{-1} \ldots k_n^{-1}, \; i>j \eqno(25) $$

\noindent
and the conjugate of (25) with respect to the antiinvolution
introduced previously.

In conclusion we should come back to  para-Bose statistics.
We have seen that at $q \rightarrow 1$ the relations (19), (20),
(21) reduce to (3), (4), (5) and, hence to the equations (1),
defining the pB operators. Therefore the operators
$A_1^{\pm},\ldots,A_n^{\pm}$ are deformed para-Bose operators.

The author is thankful to Prof. Abdus Salam for the kind
hospitality at the International Center for Theoretical Physics,
Trieste. He is grateful to
Prof. Vanden Berghe for the possibility to visit the
Department of Applied Mathematics and Computer Science at the
University of Ghent, where the paper was completed.
It is a pleasure to thank Dr. J. Van der Jeugt and Dr. N. I.
Stoilova for stimulating discussions.

The research was supported through  contract $\Phi - 215$ of
the Committee of Science of Bulgaria.

\vfill \eject

\noindent
{\bf References}

\vskip 12pt

\settabs\+[11] & I. Patera, T. D. Palev, Theoretical interpretation of the
   experiments on the elastic \cr
   %sample line,  see p. 232 of the Texbook.

\+[1] & Green H S 1953 {\it Phys.Rev.} {\bf 90}  270 \cr

\+[2] & Kamefuchi S. and Takahashi Y. 1960 {\it Nucl.Phys.}
        {\bf 36} 177 \cr

\+    & Ryan C and Sudarshan E C G 1963 {\it Nucl.Phys.}
        {\bf 47} 207 \cr

\+[3] & Omote M, Ohnuki Y  and Kamefuchi S 1976 {\it Prog.Theor.Phys.}
        {\bf 56} 1948 \cr

\+[4] & Kac V G 1978 {\it Lect.Notes Math.} {\bf 626} 597  \cr

\+[5] & Ganchev A and Palev T D 1980 {\it J.Math.Phys.}
      {\bf 21} 797 \cr

\+[6] & Wigner E P 1950 {\it Phys. Rev.} {\bf 77} 711 \cr

\+[7] & Palev T D 1982 {\it J. Math. Phys.} {\bf 23} 1778 \cr

\+[8] & Ohnuki Y and Kamefuchi 1982 {\it Quantum Field Theory
        and Parastatistics} \cr
\+    & (Univ. of Tokyo Press, Springer-Verlag, Berlin) \cr

\+[9] & Palev T D 1979 {\it Czech. Journ. Phys.} {\bf B29} 91;
        1980 {\it Rep. Math. Phys.} {\bf 18} 117, 129; \cr
\+    & 1990 Lie
        superalgebras, infinite-dimensional algebras and quantum
        statistics: \cr
\+    & 1992 {\it Rep. Math. Phys.}{\bf 31} 141 \cr

\+[10]& Floreanini R, Spiridonov V P  and Vinet L
        1991 {\it Comm.Math.Phys.} {\bf 137} 149 \cr

\+[11]& Khoroshkin S M  and Tolstoy  V N  1991 {\it Comm.Math.Phys}
        {\bf 141} 599 \cr

\+[12]& Celeghini E, Palev T D and Tarlini M 1991
        {\it Mod. Phys. Lett. B } {\bf 5} 187 \cr

\+[13]& Palev T D and N I Stoilova 1993 On a possible algebra
        morphism of $U_q[osp(1/2n)]$ onto \cr
\+    & the deformed oscillator
        algebra $W_q(n)$, {\it Preprint} ICTP IC/93/54,
        {\it Lett. Math. Phys.} (to appear) \cr

\+[14]& Flato M, Hadjiivanov L K  and Todorov I T   1993 Quantum
        Deformations of Singletons and of Free \cr
 \+   & Zero-mass Fields {\it Foundations of Physics}
        the volume dedicated to A. O. Barut (to appear)\cr

\+[15]& Floreanini R and Vinet L 1990 {\it J. Phys. A} {\bf 23}
        L1019 \cr
\+[16]& Greenberg O W in 1993 {\it Workshop on Harmonic
        Oscillators} \cr
\+    & (NASA Conference Publications 3197,
        Edited by Han D, Kim Y S and Zachary W W)  \cr

\+[17]& Palev T D 1993 A superalgebra morphism of
        $U_q[osp(1/2n)]$  onto the deformed oscillator \cr
\+    & superalgebra $W_q(n)$, {\it Preprint}
        University of Ghent TWI-93-17,
        {\it Lett. Math. Phys.} (to appear) \cr

\+[18]& Biedenharn  L C  1989 {\it J.Phys. A} {\bf 22}  L873 \cr

\noindent
\+[19]& Macfarlane  A J  1989 {\it J.Phys. A} {\bf 22}  4581 \cr

\noindent
\+[20]& Sun  C P  and Fu  H C  1989 {\it J.Phys. A} {\bf 22} L983 \cr

\+[21]& Pusz W and Woronowicz S L 1989
        {\it Rep. Math. Phys.}{\bf 27} 231 \cr

\+[22]& Hadjiivanov L K 1993 Quantum deformations of Bose
        parastatistics {\it Preprint} ESI 20, Vienna \cr

\end